\documentclass[12pt]{iopart}
\usepackage{iopams}
\usepackage{hyperref}
\usepackage{color}
\usepackage[a4paper]{anysize}

\definecolor{darkred}{rgb}{0.5,0.0,0.0}
\definecolor{darkgreen}{rgb}{0.0,0.5,0.0}
\definecolor{darkblue}{rgb}{0.0,0.0,0.5}
\hypersetup{colorlinks=true,
citecolor=darkred,
linkcolor=darkgreen,
urlcolor=darkblue}
\marginsize{3cm}{2.5cm}{2.0cm}{4.0cm}

\usepackage{graphicx}
\usepackage{bm}
\usepackage{bbm}
\usepackage[applemac]{inputenc}
\usepackage{iopams}

\begin{document}

\title[Correlations and entanglement in a semiconductor hybrid circuit-QED system]{Non-equilibrium correlations and entanglement in a semiconductor hybrid circuit-QED system}

\author{L.~D. Contreras-Pulido$^1$, C. Emary$^{2,3}$, T. Brandes$^3$, Ram\'on Aguado$^4$}
\address{$^1$Institut f\"ur Theoretische Physik, Albert-Einstein Allee 11, Universit\"at Ulm, D-89069 Ulm, Germany\\$^2$Department of Physics and Mathematics, University of Hull, Kingston-upon-Hull, HU6 7RX, United Kingdom\\$^3$Institut f\"ur Theoretische Physik, Hardenbergstrasse 36, TU Berlin, D-10623 Berlin, Germany\\$^4$Instituto de Ciencia de Materiales de Madrid (ICMM), Consejo Superior de Investigaciones Cient\'ificas (CSIC), Sor Juana In\'es de la Cruz 3, 28049 Madrid, Spain}
\ead{debora.contreras@uni-ulm.de}

\begin{abstract}

We present a theoretical study of a hybrid circuit-quantum electrodynamics system composed of two semiconducting charge-qubits confined in a microwave resonator. The qubits are defined in terms of the charge states of two spatially separated double quantum dots (DQDs) which are coupled to the same photon mode in the microwave resonator. We analyze a transport setup where each DQD is attached to electronic reservoirs and biased out-of-equilibrium by a large voltage, and study how electron transport across each DQD is modified by the coupling to the common resonator. In particular, we show that the inelastic current through each DQD reflects an indirect qubit-qubit interaction mediated by off-resonant photons in the microwave resonator. As a result of this interaction, both charge qubits stay entangled in the steady (dissipative) state. Finite shot noise cross-correlations between currents across distant DQDs are another manifestation of this nontrivial steady-state entanglement.

\end{abstract}

\maketitle

\section{Introduction}

Recent technological progress has made it possible to coherently couple superconducting qubits to microwave photons on a superconducting chip \cite{wallraff04}. This so-called circuit quantum electrodynamics (circuit-QED) \cite{blais04} has paved the way for new research directions beyond standard cavity QED systems \cite{nori05,girvin08,nori11}. Apart from the high degree of tunability in circuit-QED, most of the novelty comes from the fact that the coupling between qubits and microwave photons can reach values well above the ones between natural atoms and photons in optical cavities \cite{deppe10}.

An interesting alternative to the above ideas is to use hybrid circuit-QED \cite{nori13} with qubits defined in semiconducting quantum dots (QDs) \cite{childress04,loss08,cottet10,nori12,jin12}. Such a concept has been recently experimentally implemented \cite{frey11,petta12,delbecq12,frey12,frey12b,toida12,delbecq13}. In these hybrid structures, the semiconducting QDs are typically coupled to normal electronic reservoirs such that electronic transport may be used to characterize/modify the properties of the circuit-QED system. Although this possibility had remained largely unexplored, except for some works analyzing the transport-induced lasing states in the resonator \cite{nori09,jin11,cottet12,brandes13,xu13}, this kind of setups are now attracting growing theoretical attention \cite{samuelsson13,lambert13}.

In this context, we here analyse how the coupling to a common photon mode generates entanglement between distant charge qubits realized in double quantum dots (DQDs) and how this entanglement manifests in the transport properties of the system. In particular, we present a detailed analysis of how the electron currents across each DQD are modified due to the interaction with the photons in the circuit. The coupling of each DQD to a common microwave resonator generates an indirect coupling between DQDs which gives rise to positive shot noise cross-correlations between \emph{distant} currents across them. We analyse this physics in terms of an effective model and show that off-resonant photons are responsible for the induced indirect coupling. Moreover, we demonstrate that both charge qubits are entangled in the steady (dissipative) state due to this resonator-induced coupling. In \cite{samuelsson13}, Bergenfeldt and Samuelsson have studied the effect that non-local interaction between two DQDs resonantly coupled to the oscillator has on finite bias voltage transport properties (which are prone to finite temperature effects in the electronic reservoirs). In contrast, we here focus on a different operating regime where the non-local interaction is induced off-resonance and transport occurs at very large voltages. In this large-voltage regime, the results are essentially independent on the electronic reservoir temperature and are valid at arbitrary couplings to the reservoirs. This large voltage regime is also analysed in \cite{lambert13},  where some overlapping results about photon-mediated transport and finite shot noise cross-correlations have been reported.

\begin{figure}
\centering
    \includegraphics[width=0.45\columnwidth]{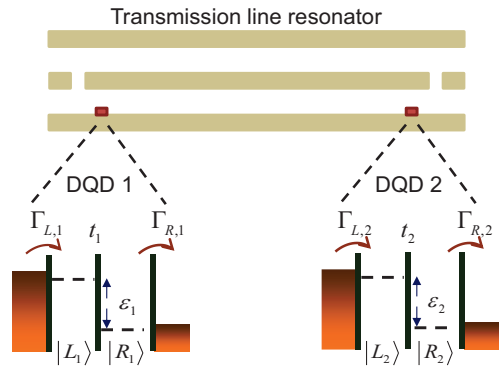}
   \caption{Schematics of the two charge qubits coupled to a transmission line resonator. An excess charge in each double dot (formed by the $L_i$ and the $R_i$ dots) defines the states of the qubit. Both qubits are attached to electronic reservoirs, via the rates $\Gamma_{L,i}$ and $\Gamma_{R,i}$, such that an electrical current pass through them. The qubits are located at the ends of the resonator in order to enhance the coupling with the electromagnetic field.}
    \label{fig_sketch}
\end{figure}

The paper is organized as follows. In section \ref{Model} we describe the model for two double quantum dots coupled to a microwave resonator as well as the master equation that governs the dynamics of this open quantum system. In section \ref{transport} we discuss the stationary transport properties (mean value of the current and shot noise) of the system. This section is divided in two parts. The first part (section  \ref{singleDQD}) reviews the case of a single double quantum dot. We then turn to the analysis of the two double quantum dot system (ssection \ref{two-DQDs}) by also calculating shot-noise cross-correlations between distant currents across each double quantum dot. In Section \ref{entanglement} we focus on the generation of qubit-qubit entanglement induced by the common coupling to a microwave photon mode, and compare it with the results obtained for the cross-correlations in the previous section. In particular, we analyse the steady-state
Bell states occupations and demonstrate that indeed cross-correlations between distant currents constitute an indicator of non-local qubit-qubit entanglement. In section \ref{asymmetric} we extend our study to the case of asymmetric couplings between each double quantum dot and the microwave resonator. Our conclusions are presented in section \ref{conclusions}.

\section{Model}\label{Model}

We consider the coupling of the charge states of two uncoupled semiconductor DQDs to an electromagnetic resonator with a high $Q$-factor, as for instance the superconducting transmission line described in the recent experiments of \cite{frey12}. We assume that the DQDs are placed at the ends of the resonator, as schematically depicted in figure~\ref{fig_sketch}. In the following, we consider that the charging energy on each DQD is the largest energy scale of the problem such that, for each individual DQD, an excess electron defines the two states of a charge qubit, $|L_{i}\rangle$ and $|R_{i}\rangle$ ($i=1,2$), see e.g. \cite{vanderwiel03}. In this basis, the Hamiltonian describing the DQDs reads
\begin{equation}
H_{\mathrm{el}}=\sum_{i}\left(\frac{1}{2}\varepsilon_{i}\sigma_{z}^{i}+t_{i}\sigma_{x}^{i}\right)
\label{Hel}
\end{equation}
where the energy detuning in each DQD is given by $\varepsilon_{i}$, $t_{i}$ is the tunnelling coupling between dots of the $i$-th DQD and $\sigma_{j}$ is the $j$-th Pauli matrix acting on the charge basis of each qubit, namely $\sigma_{z}^{i}\equiv |L_{i}\rangle\langle L_{i}|-|R_{i}\rangle\langle R_{i}|$ and $\sigma_{x}^{i}\equiv |L_{i}\rangle\langle R_{i}|+|R_{i}\rangle\langle L_{i}|$.

The transmission line resonator is modelled as a quantum harmonic oscillator $H_{\mathrm{res}}=\hbar\omega_r a^{\dagger}a$, where $a^{\dagger}$ ($a$) is the creation (annihilation) operator of photons in the resonator with frequency $\omega_r$.
The charge states of each qubit are coupled to the same mode of the resonator, such that the coupling term reads
\begin{equation}
H_{\mathrm{e-res}}=\sum_{i}\hbar g_{i}\sigma_{z}^{i}(a^{\dagger}+a).
\label{Helph}
\end{equation}
Experimentally, typical photon frequency takes values $\omega_r/2\pi\sim$1-10 GHz, whereas couplings strengths $g/2\pi\sim$10-30 MHz have been reported for a single DQD coupled to a microwave resonator \cite{frey12,frey12b}.

Finally, we consider that each DQD ($i$=1,2) is attached to electronic reservoirs, which are described by the Hamiltonian
\begin{equation}
H_{\mathrm{leads}}=\sum_{i}\sum_{k}\{\varepsilon_{k,i}^{L}c_{k_L,i}^{\dagger}c_{k_L,i}+\varepsilon_{k,i}^{R}c_{k_R,i}^{\dagger}c_{k_R,i}\}
\label{Hres}
\end{equation}
in which $c_{k_\beta,i}^{\dagger}(c_{k_\beta,i})$ is the creation (annihilation) operator of electrons in the left/right contact, $\beta\in L,R$, with energy $\varepsilon_{k,i}^{\beta}$. The coupling of each DQD to the leads reads:
\begin{equation}
H_{\mathrm{int}}=\sum_{i}\sum_{k}\{V_{k,i}^{L}c_{k,L,i}^{\dagger}d_{L,i}+h.c.+L\rightarrow R\}
\label{HSR}
\end{equation}
where $d_{L/R,i}$ $(d_{L/R,i}^{\dagger})$ creates (annihilates) an electron in the left/right QD of each DQD, and $V_{k,i}^{\beta}$ are the tunnelling matrix elements. Due to this coupling to the reservoirs, situations in which either of the two DQDs (or both) are empty need to be considered and hence the Hilbert space in the charge sector is spanned by the states $|\alpha_1,\alpha_2\rangle$, with $\alpha=L, R, 0$. This transport model can be easily extended to a system consisting of several qubits, see e.g. \cite{lambert09}, and is the single-mode version of previous studies focusing on bath-mediated interactions \cite{phbath}.

The total Hamiltonian of the system is given by $H_{\mathrm{tot}}=H_{\mathrm{el}}+H_{\mathrm{res}}+H_{\mathrm{e-res}}+H_{\mathrm{leads}}+H_{\mathrm{int}}$.  The dynamics of the resonator and the DQDs is described by the master equation for the reduced density matrix $\rho(t)$ obtained after tracing out the reservoirs degrees of freedom and applying a Born-Markov approximation with respect to the Hamiltonian $H_{\mathrm{int}}$ \cite{breuer}. In the Schr\"odinger picture the master equation reads $\dot{\rho}(t)=\mathcal{L}\rho $ with the Liouvillian:
\begin{eqnarray}
&&\mathcal{L}\rho=-i\left[H_{\mathrm{el}}+H_{\mathrm{res}}+H_{\mathrm{e-res}},\rho(t)\right]\nonumber\\
&&-\sum_{i}\frac{\Gamma_{i}^{L}}{2}\left(d_{L,i}d_{L,i}^{\dagger}\rho(t)-2d_{L,i}^{\dagger}\rho(t)d_{L,i}+\rho(t)d_{L,i}d_{L,i}^{\dagger}\right)\nonumber \\
&&-\sum_{i}\frac{\Gamma_{i}^{R}}{2}\left(d_{R,i}^{\dagger}d_{R,i}\rho(t)-2d_{R,i}\rho(t)d_{R,i}^{\dagger}+\rho(t)d_{R,i}^{\dagger}d_{R,i}\right)\nonumber\\
&&-\frac{\kappa}{2}\left(2a\rho(t)a^{\dagger}-a^{\dagger}a\rho(t)+\rho(t)a^{\dagger}a\right)
\label{RDM}
\end{eqnarray}

with the tunnelling rates to reservoirs $\Gamma_{i}^{\beta}=2\pi\sum_{k,i}|V_{k,i}^{\beta}|^{2}\delta(\varepsilon_{i,\beta}-\varepsilon_{k,i,\beta})$ and where we considered the limit of infinite source-drain voltage, $\mu_L\rightarrow\infty$ and $\mu_R\rightarrow -\infty$ (such that the Fermi functions in the reservoirs become $f_L=1$ and $f_R=0$). In this limit, the Born-Markov approximation with respect to the coupling to reservoirs is essentially exact and, more importantly, the physics no longer depends on the temperature of the electronic reservoirs \cite{stoof96,brandesrep}. A finite zero-temperature damping in the cavity, with rate $\kappa$ \cite{milburn}, has also been taken into account by including the last Lindblad term in equation~(\ref{RDM}).

We are interested on the generation of qubit-qubit entanglement and on the transport properties in the stationary state, $\rho^{\mathrm{stat}}$. This can be obtained from equation~(\ref{RDM}) as $\dot{\rho}(t)=\mathcal{L}\rho^{\mathrm{stat}}=0$ such that the Liouvillian $\mathcal{L}$ has a zero eigenvalue with right eigenvector denoted as $|0\rangle\rangle=\rho^{\mathrm{stat}}$. The corresponding left eigenvector is $\langle\langle\tilde{0}|$ such that the probability conservation reads $\langle\langle\tilde{0}|0\rangle\rangle=\mathrm{Tr}[\hat{1}\rho^{\mathrm{stat}}]=1$. Using this language, the average of any operator $\hat{A}$ acting on the qubits-resonator system reads $\langle\hat{A}\rangle=\mathrm{Tr}[\hat{A}\rho^{\mathrm{stat}}]=\langle\langle\tilde{0}|\hat{A}|0\rangle\rangle=\langle\langle\hat{A}\rangle\rangle$.

The set of equations for the elements of the density matrix $\rho_{nm}(t)$, in the basis given by  the direct product of the electronic states and the oscillator Fock states $|\alpha_1,\alpha_2\rangle\otimes|n\rangle$ (with $n=0,1,2,...$), is solved numerically by truncating up to a maximum number of photon states $n=N_{\mathrm{max}}$.\footnote{With the order of magnitude of the parameters used here, $N_{\mathrm{max}}=6$ is sufficient to achieve numerical convergence.} We take the order of magnitude of the parameters from the recent experiments reporting circuit-QED devices with semiconducting QDs \cite{frey12,frey12b}. Even though we focus here on this moderate coupling regime ${g/\omega_r}\sim10^{-2}$, we note in passing that our numerical scheme allows in principle to include stronger couplings, such as the ones already achieved in circuit-QED architectures with superconducting qubits \cite{deppe10,casanova10}.

\section{Stationary transport properties: Current, shot noise and current correlations}\label{transport}

We expect that the indirect, non-local two-qubit interaction induced by the coupling to a common resonator mode can be revealed in transport through either DQD. As previously mentioned, we restrict ourselves to the Coulomb blockade regime in the infinite bias voltage limit.

In this case of unidirectional transport, the total current across the DQD$i$ is described by the operator $\mathcal{I}_i\rho=e\Gamma_{R,i}d_{R,i}\rho d^{\dagger}_{R,i}$, and the corresponding steady-state expectation value reads $I_i=\langle\langle\tilde{0}|\mathcal{I}_i|0\rangle\rangle=Tr[\mathcal{I}_i\rho^{stat}]$.

We also analyse the non-equilibrium quantum noise, resulting from the temporal fluctuations of the current, by means of the current-current correlation function $\langle\Delta I_i(\tau),\Delta I_j(0)\rangle$, with $\Delta I_i(t)=I_i(t)-\langle I_i\rangle$. The Fourier transform of such correlation function defines the power spectral density of shot noise:
\begin{equation}
S_{ij}(\omega)=2\int_{-\infty}^{\infty}d\tau e^{i\omega\tau}\langle\{\Delta I_i(\tau),\Delta I_j(0)\}\rangle
\label{shotnoise}
\end{equation}
It has been shown that this finite-frequency power spectral density contains a great deal of information about internal dynamics of the system \cite{sfreq}. Nevertheless, we here restrict the analysis to zero frequencies for simplicity. Due to the possibility of individual control and manipulation of the QDs, in particular we focus on the cross-correlations which, as we shall show, exhibit features related with the qubit-qubit effective interaction induced by the common coupling to the resonator. Additional interest in studying shot noise and cross-correlations reside in theoretical proposals which make use of current correlations to study and detect entanglement in mesoscopic systems \cite{burkard00,lesovik01,plastina01,lesovik02,beenakker03,samuelsson03,emary09,belzig11}.

In practice, the shot noise at zero-frequency is calculated in terms of the inverse of the part of the Liouvillian that is non-singular at zero-frequency (or pseudo-inverse), $R=\mathcal{Q}\mathcal{L}^{-1}\mathcal{Q}$ (with $\mathcal{Q}=1-|0\rangle\rangle\langle\langle\tilde{0}|$), see e.g. \cite{flindt04,lambert07,emary09b}. The diagonal part of the noise reads $S_{ii}(0)=2(\langle\langle \mathcal{I}_{i}\rangle\rangle-2\langle\langle \mathcal{I}_{i}R\mathcal{I}_{i}\rangle\rangle)$, with $i=1,2$, whereas the off-diagonal noise cross-correlations read $S_{12}(0)=S_{21}(0)=-2\left(\langle\langle \mathcal{I}_{1}R\mathcal{I}_{2}\rangle\rangle+\langle\langle \mathcal{I}_{2}R\mathcal{I}_{1}\rangle\rangle\right)$. Note that any finite off-diagonal noise in this setup indicates correlations between \emph{distant} currents across each DQD.

In what follows we present our noise results in the form of Fano factors, defined as $F_{ij}=S_{ij}(0)/(2e\sqrt{I_iI_j})$, which quantifies deviations from the Poissonian noise originated by uncorrelated carriers. In particular, super-Poissonian noise $(F>1)$ is related to a bunching behaviour of the carriers whereas sub-Poissonian noise $(F<1)$ signals anti-bunching. For the relation between (anti)bunching and the Fano factor in electronic transport, see \cite{emary12}.

\subsection{A single DQD coupled to the resonator}\label{singleDQD}

To set the stage for our study, we begin by analysing the case of a single DQD coupled to a microwave resonator. The physics here is that of inelastic transport through a two-level system, a problem which has received a lot of attention in various contexts \cite{brandes99,ramon00,ramon04,brandesrep,lambert03,dong05,lambert08,harvery08,bruder09,granger12}. In the frame of circuit-QED with semiconducting qubits, the problem has been theoretically studied in \cite{jin11,brandes13} mainly with focus on lasing.

\begin{figure}[t]
\centering
   \includegraphics[width=0.85\columnwidth]{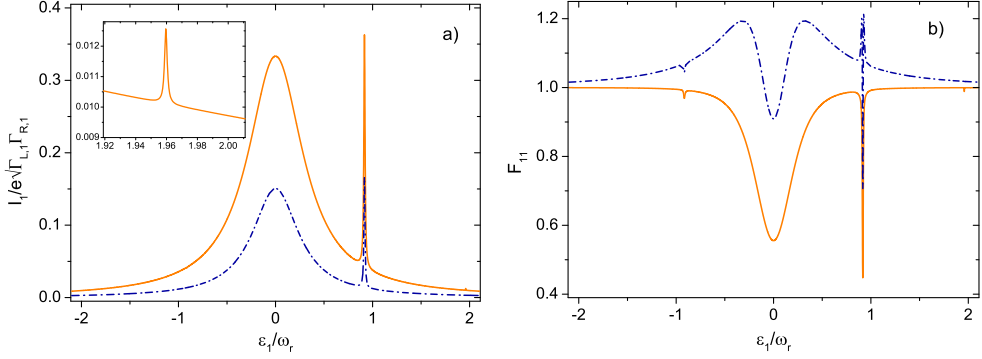}
   \caption{Results for a) stationary current and b) Fano factor for a single-qubit coupled to a transmission line resonator. The solid line corresponds to equal tunnelling rates to the reservoirs $(\Gamma_{L,1}=\Gamma_{R,1}=10^{-3})$ and the dashed line to asymmetric rates $(\Gamma_{L,1}=0.01, \Gamma_{R,1}=10^{-3})$. Note that the inelastic peaks appear at values of $\varepsilon_1$ corresponding to the resonance condition $\Omega_1=n\hbar\omega_r$. Inset: zoom of the current peak at $\Omega_1=2\hbar\omega_r$ for symmetric rates. The rest of the parameters (in units of $\omega_r$) are: $t_1=0.2$, $g_1=0.008$ and $\kappa=5\times10^{-4}$.}
\label{fig_1qbs}
\end{figure}

In figure~\ref{fig_1qbs}a) we show the current in the DQD as a function of its level detuning $\varepsilon_1$ (all the parameters are expressed in units of the resonator frequency $\omega_r$). As expected, there is an elastic peak in the current around $\varepsilon_1=0$ which corresponds to resonant tunnelling across the DQD. Here, the electronic transport occurs by the tunnel coupling with the reservoirs, which we assumed to be the same for both leads $\Gamma_{L,1}=\Gamma_{R,1}$. The height and width of the elastic peak is in agreement with the well known analytical expression for the current through a DQD \cite{nazarov93,stoof96}. For finite detuning (i.e., with the electronic levels of the DQD far from resonance) the current is suppressed except at values of $\varepsilon_1$ corresponding to a resonance condition at which the frequency of the qubit $\Omega_1\equiv\sqrt{\varepsilon_1^2+4t_1^2}$ equals the frequency of the resonator $\omega_r$. This feature corresponds to inelastic processes in which the tunnelling of an electron between the left and right dots of the qubit excites the state of the resonator. This behaviour is in qualitative agreement with the theoretical results of Jin \textit{et al}., \cite{jin11} who studied lasing in a DQD-based circuit-QED system (the main idea being that transport of electrons through the artificial two-level system can lead
to a population inversion and induce a lasing state in the microwave resonator, as studied for superconducting qubit-based architectures, see e.g., \cite{nori13}. Indeed lasing in a Cooper pair box coupled to a superconducting resonator was experimentally demonstrated in \cite{JQP}). Although we are not interested on analysing the specific lasing conditions, the underlying mechanism giving rise to the inelastic peak of the current is the same. For large enough electron-boson coupling $g_1$, additional resonances at $\Omega_1\approx n\hbar\omega_r$ appear. An example for $n=2$ is shown in the inset in figure~\ref{fig_1qbs}a).

The physics above is very similar to the one of spontaneous emission of a DQD coupled to a dissipative bath of phonons \footnote{As well as the physics of on-chip noise detection using two-level systems \cite{ramon00,noisedet,granger12}.}. In fact, a spontaneous emission background will always coexist with the photon emission peaks we just discussed. The reason is simple: the DQD is never truly isolated from the environment and even near zero temperature there is a finite current for $\varepsilon_1>0$ due to quantum fluctuations. This spontaneous emission contribution to the inelastic current due to vacuum fluctuations was first demonstrated experimentally in \cite{fujisawa98}.

More specifically, one can estimate the role that dissipative effects have on our scheme by including in the model a dipolar coupling to a bosonic bath (very much like the coupling in Eq.~(\ref{Helph}), but replacing the single mode cavity by a full bath of bosons, namely $H_{\mathrm{e-bath}}=\sum_{\bf q} \frac{g_{\bf q}}{2}(b^{\dagger}_{\bf q}+b_{\bf q})\sigma_{z}$). Within the Born-Markov approximation this leads to an energy relaxation rate of the form (for details, see e.g. \cite{brandesrep}):
$\gamma_1\equiv 2\pi \frac{t^2_1}{\Omega^2_1} J(\frac{\Omega_1}{\hbar}) \coth
\left(\frac{\Omega_1}{2k_BT}\right)$, where the effects
of the dissipative bosonic bath are fully encapsulated in the spectral density
$J({\omega})\equiv \sum_{\bf q} |g_{\bf q} |^2\delta(\omega-\omega_{\bf q})$. In GaAs-AlGaAs DQDs the energy relaxation is primarily dominated by the emission of piezoelectric acoustic phonons which in the simplest approximation (bulk limit and vanishing longitudinal speed of sound) can be described by an Ohmic bath $J({\omega})=2\alpha\omega e^{-\omega/\omega_c}$, where $\omega_c$ is a high frequency cutoff \footnote{If one considers a more realistic bath of piezoacoustic phonons, the spectral function reads $J(\omega)=2{\alpha}{\omega}\left[1-{\omega_d}/{\omega}\sin\left({\omega}/{\omega_d}\right)\right]
e^{-\omega/\omega_c}$, with $\omega_d$ depending on geometry (for details, see Ref. \cite{brandesrep})}. The total DQD decoherence rate is given by $\gamma=\gamma_1/2+\Gamma_R/2+\gamma_\phi$, where $\gamma_\phi$ is the pure dephasing rate which for an Ohmic bath reads $\gamma_\phi=2\pi\alpha(\frac{\varepsilon_1}{\Omega_1})^2k_BT$. The advantage of such simple parametrization of the bath is that it allows to estimate the coupling parameter $\alpha$ by just substituting DQD parameters from a given experiment.
For example, in the experimental work by Frey \textit{et al} \cite{frey12}, typical relaxation rates for charge qubits in the large detuning regime $\varepsilon_1>t_1$ are $\frac{\gamma_1}{2\pi}\approx 100$ MHz, while dephasing rates range from $\frac{\gamma_\phi}{2\pi}\approx 1-3$ GHz. This qubit dephasing rate is significantly larger than the coupling strength $\frac{g}{2\pi}\approx 50$ MHz, so a vacuum Rabi mode splitting, implying a fully quantum coherent interaction
between the DQD and the cavity, is not observed (subsequent experiments \cite{toida12} claimed much
smaller dephasing rates and hence a strong cavity-qubit coupling regime, however the analysis
of these experiments has been recently questioned in \cite{deph}). Instead, the observed frequency shift and linewidth broadening of the resonator in the experiments are consistent with a dipole coupling of several tens of MHz to the resonator. Since the effects we are discussing here do not involve the stringent condition of working with full hybrid cavity-qubit states, namely a strong coupling limit, we expect that qubit decoherence is not a major obstacle for the physics we shall be discussing in the following \footnote{A systematic study of decoherence effects on transport and noise in a circuit QED system based on DQDs can be found in \cite{jin11}. The main effect of decoherence is that transport resonances involving photons just become broadened, which supports our arguments.}.

The corresponding Fano factor $F_{11}$, shown in figure~\ref{fig_1qbs}b), exhibits a dip around $\varepsilon_1=0$. There, interdot tunnelling delocalizes the charge which, combined with the strong Coulomb blockade, reduces the noise and gives sub-Possonian Fano factor, $F_{11}<1$ \cite{ramon04,choi03}. As the level detuning $\varepsilon_1$ increases, the charge becomes localized, say in the left dot for $\varepsilon_1>0$, and hence Poissonian noise from a single barrier (the one parametrized by $\Gamma_{L}$) is obtained. This is so until the resonance conditions $\Omega_1=n\hbar\omega_r$ are reached, where the noise is reduced again yielding $F_{11}<1$. This sub-Poissonian value at resonance with the photon mode reveals that the charge is transferred across the DQD with the simultaneous excitation of the resonator. The same kind of result is obtained for emission into a full bath of bosons \cite{ramon04}.

Note also the small resonant feature in the region $\varepsilon_1<0$. Even though in this configuration the extra charge is mainly localized in the left dot, there is a small probability of populating the right dot (and subsequently tunnel out from the right barrier). From the point of view of the qubit, this means that there is a small probability of populating the excited state and hence to emit photons. This can be easily seen if we write the qubit-photon interaction in the qubit eigenbasis $|e\rangle=\cos\frac{\theta}{2}|L\rangle+\sin\frac{\theta}{2}|R\rangle$ and $|g\rangle=-\sin\frac{\theta}{2}|L\rangle+\cos\frac{\theta}{2}|R\rangle$, with $\theta=\arctan(\frac{2t_1}{\varepsilon_1})$ being the angle that characterizes mixing in the charge subspace: $H_{\mathrm{e-res}}=g_1(\cos\theta\tau_z+\sin\theta\tau_x)(a^{\dagger}+a)$, with $\tau_z=|e\rangle\langle e|-|g\rangle\langle g|$ and $\tau_x=|e\rangle\langle g|+|g\rangle\langle e|$. We have checked that photon emission at $\varepsilon_1\approx-1$ is small but finite (the photon occupation has a resonance around this detuning and increases from zero to $\langle n\rangle\approx 10^{-3}$, not shown), as a result of $|e\rangle\rightarrow |g\rangle$ relaxation processes. Dynamically, these rare events, where the qubit is excited for negative detuning such that photon emission is possible, contribute to the noise which shows a feature at $\Omega_{1}=\hbar\omega_r$ with $\varepsilon_1<0$. On the contrary, they do not change significantly the average current, demonstrating the superior sensitivity that noise has.

\begin{figure}
\centering
    \includegraphics[width=0.85\columnwidth]{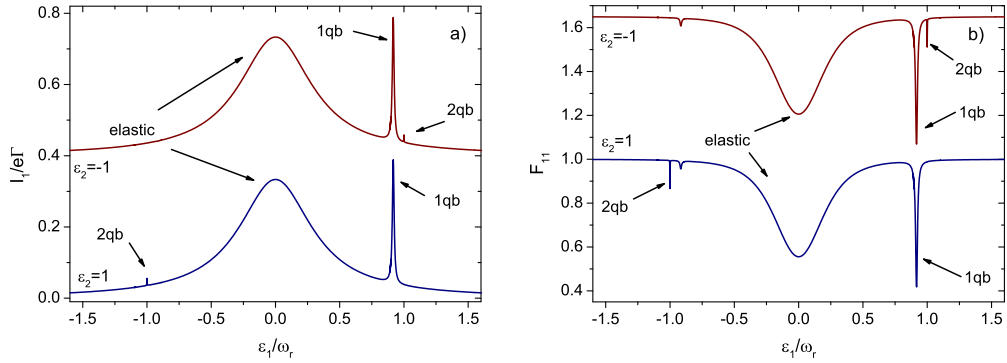}
   \caption{a) Steady state current and b) Fano factor in the first DQD as a function of its level position $\varepsilon_{1}$ for two different configurations of the second qubit: $\varepsilon_{2}=-1$ and $\varepsilon_{2}=1$. The two examples have been vertically shifted (with an offset of 0.4 and 0.6 respectively) for the sake of clarity. Parameters (in units of $\omega_r$): $g_{1}=g_{2}=g=0.008$, $t_{1}=t_{2}=t=0.2$, $\Gamma=10^{-3}$ and $\kappa=10^{-3}$.}
\label{fig_I2qb}
\end{figure}

The effect on the transport properties of asymmetric tunnelling rates is also shown in figure~\ref{fig_1qbs}, where we considered that $\Gamma_{L,1}>\Gamma_{R,1}$. The current, figure~\ref{fig_1qbs}a), exhibits the same qualitative behaviour than the case with equal rates, with an elastic peak around $\varepsilon=0$ and satellite peaks at the resonances qubit-resonator.
On the contrary, the Fano factor changes completely for asymmetric rates, figure~\ref{fig_1qbs}b). In this case,  $F_{11}$ presents a double peak structure in the region $\varepsilon=0$, with the maximum of the peaks corresponding to super-Poissonian noise. This well-known effect can be understood from the analytical expression of the Fano factor \cite{elattari02} and, in particular, is originated from the smaller coupling to the drain reservoir which, ultimately, makes the Coulomb interaction more effective and gives rise to bunching in transport with $F_{11}>1$. The same kind of bunching behaviour is observed for the Fano factor at the qubit-photon resonances. It is interesting to compare this $F_{11}>1$ at the one photon resonance with the result for a full bosonic bath which always results in sub-Poissonian noise \cite{ramon04,kiesslich07}. Hence super-Poissonian noise results from the qubit-photon coherent interaction. This result is also along the lines of \cite{kiesslich07}, where the authors demonstrate that the bunching effect cannot be obtained from a picture without qubit coherences. In the context of lasing, this sort of super-Poissonian noise has been related to squeezing of the resonator state \cite{jin12b}.

\subsection{Two DQDs coupled to the transmission line resonator}\label{two-DQDs}

We turn now to our original model in which two DQDs are coupled to the same photon mode of the microwave resonator, but uncoupled to each other. For simplicity, we consider first the same intra-dot tunnel couplings $t_{i}=t$ and equal electron-photon coupling $g_{i}=g$. It is assumed that the tunnelling rates to left and right reservoirs are equal and also equivalent for both DQDs i.e., $\Gamma_{L,i}=\Gamma_{R,i}=\Gamma$, unless otherwise stated. As in the case for a single DQD, all the parameters are given in terms of the bare frequency of the microwave resonator $\omega_r$.

Results for the mean value of the stationary current across the first DQD, $I_1$, as function of its level detuning $\varepsilon_1$, while keeping the second DQD in a fixed level structure, are presented in figure~\ref{fig_I2qb}a). Similarly to the case for a single DQD, there is an elastic peak in $I_1$ around $\varepsilon_{1}=0$. A second, inelastic peak appears in the region where this qubit enters in resonance with the photon, $\Omega_1\approx\hbar\omega_r$ revealing that this effect is entirely due to the coupling of this DQD with the resonator and thus will appear irrespective of the presence of the second DQD. We refer to this feature as the \textit{one-qubit (1qb) peak}.

Interestingly, an additional peak in $I_1$ is observed in the emission part $\varepsilon_{1}>0$ for the case $\varepsilon_{2}=-1$, and in the absorption part $\varepsilon_{1}<0$ with $\varepsilon_{2}=1$. The third peak arises when both DQDs are brought in resonance with each other, $\Omega_{1}=\Omega_{2}$, with opposite detuning, $\varepsilon_1=-\varepsilon_2$, but slightly out of resonance with the photon mode, $\Omega_{1}=\Omega_{2}\neq\hbar\omega_r$. It is a result of an indirect qubit-qubit interaction induced by the common coupling to the microwave resonator and therefore we refer to it as the \textit{two-qubits (2qb) peak}. The fact that this resonance appears at an energy larger than the frequency $\omega_r$ reveals that the effective interaction is obtained via virtual photons: when both qubits are in resonance, the excitation in one of the DQDs is transferred to the other by virtually becoming a photon in the microwave resonator. 
Similar physics has been demonstrated experimentally in circuit-QED systems with superconducting qubits, see \cite{majer07}.

In order to have a better understanding of the induced qubit-qubit interaction, we derive an effective Hamiltonian valid in the regime $|\Omega_i-\omega_r|>g$ where the 2qb-features appear.
In this regime, with both qubits on resonance with each other but off-resonance with the mode, such that the resonator remains essentially in its ground-state with the interaction between the qubits
mediated by virtual photons. We can describe this situation with an effective Hamiltonian \cite{Soliverez1981} that acts in the sub-space spanned by the states $|s_1,s_2,0\rangle$ where the resonator is empty and $s_i=L_i,R_i$ describe the qubit states. Performing second-order perturbation theory for the action of the Hamiltonian $H_2=H_{\mathrm{el}}+H_{\mathrm{res}}+H_{\mathrm{e-res}}$
within this sector and restricting excitations to single photon states, $|s'_1,s'_2,1\rangle$, we obtain (see the appendix)
\begin{equation}
  H_{\mathrm{eff}}=\sum_{i}\left(\frac{1}{2}\varepsilon_{i}\sigma_{z}^{i}+t'_{i,\mathrm{eff}}\sigma_{x}^{i}\right)+J'_{z}\sigma_{z}^{1}\sigma_{z}^{2}-\sum_{i\neq j}J'_{xz,ij}\sigma_{z}^{i}\sigma_{x}^{j}
  \label{Heff}
  ,
\end{equation}
where a constant term has been neglected. The effective Hamiltonian of equation~(\ref{Heff}) explicitly shows that the interaction of the qubits with a common photon mode translates into a shift of their frequencies, through the renormalized tunnelling amplitude
\begin{eqnarray}
  t'_{i,\mathrm{eff}}&=&t_i\left[1+\frac{g_i^2}{\Omega_i}\left(\frac{1}{\Omega_i-\omega_r}\right)\right]
  ,
\end{eqnarray}
as well as two types of qubit-qubit interaction. The first one is Ising-like with effective exchange constant
\begin{eqnarray}
  J'_{z}&=&\sum_{i}\frac{2g_{1}g_{2}t_i^2}{\Omega_i^2(\Omega_i-\omega_r)}
  ,
  \label{JeffZ}
\end{eqnarray}
whereas the second one is an $XZ$ exchange interaction with a coupling strength
\begin{eqnarray}
  J'_{xz,ij}&=&\frac{g_{i}g_{j}\varepsilon_jt_j}{\Omega_j^2}
  \left(\frac{1}   {\Omega_j-\omega_r}\right)
  \label{JeffXZ}
  .
\end{eqnarray}
In these expressions for the effective couplings we have made the further assumptions $|\Omega_i-\omega| \ll \omega$ and  $|\Omega_i-\omega| \ll |\Omega_i+\omega|$ consistent with the dispersive limit and the rotating-wave approximation\footnote{Note that we have effectively removed the photons from the problem, this is the reason why the Hamiltonian is not in the standard dispersive form.}.

\begin{figure}
\centering
   \includegraphics[width=0.5\columnwidth]{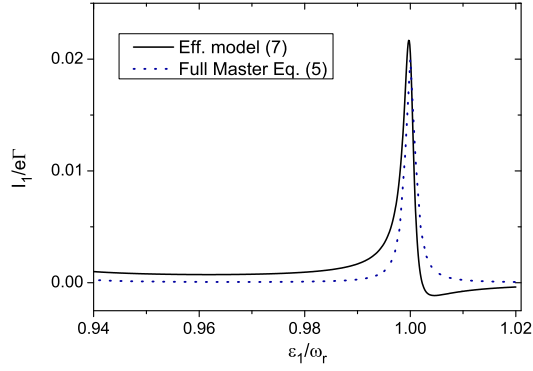}
   \caption{Comparison of the results for the steady state current in the first DQD as a function of its level position $\varepsilon_{1}$, obtained with the full Master Equation (\ref{RDM}) and with the model described by the effective Hamiltonian~(\ref{Heff}). Parameters: $\varepsilon_{2}=-1$, $g=0.008$, $t=0.2$, $\Gamma=10^{-3}$ and $\kappa=10^{-3}$.}
\label{fig_comparision}
\end{figure}

The interaction terms in equation~(\ref{Heff}) capture quite well the 2qb transport features as shown in figure~\ref{fig_comparision}, where we plot a comparison of the current calculated with an effective master equation obtained from the model~(\ref{Heff}) against the one obtained with the full master equation given by (\ref{RDM}), around the two-qubits resonance condition $\Omega_1=\Omega_2$. There, it can be noticed that the effective model reproduces the width and height of the 2qb peak. Outside this qubit-qubit
resonance condition the effective Hamiltonian is no longer valid and, therefore, cannot describe
transport in the full regime of level detunings\footnote{The results presented in the following are obtained with the full master equation (\ref{RDM}), while the effective Hamiltonian (\ref{Heff}) will be used to understand the transport features appearing at the qubit-qubit resonance $\Omega_1=\Omega_2$.}.

Once we have shown that the 2qb feature comes indeed from a resonator-induced interaction between both charge qubits, we describe how the non-local character of this interaction can be easily explored.

\begin{figure}
\centering
    \includegraphics[width=0.85\columnwidth]{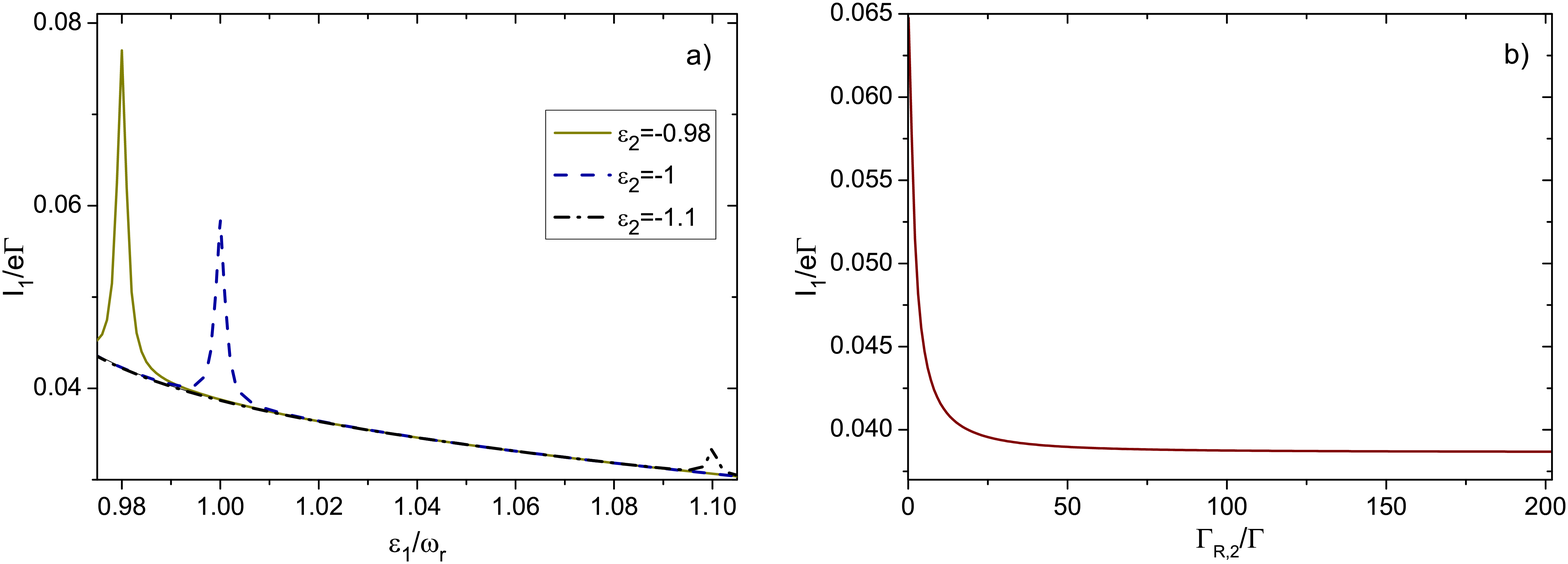}
   \caption{Current on the first DQD around the qubit-qubit resonance a) as a function of $\varepsilon_{1}$ for different values of $\varepsilon_2$ (Rest of the parameters as those used in figure~\ref{fig_I2qb});  b) as a function of $\Gamma_{R,2}$ (parameters in units of $\omega_r$: $\varepsilon_1=-\varepsilon_2=1$, $t=0.2$, $\Gamma_{L,1}=\Gamma_{R,1}=\Gamma_{L,2}=10^{-3}$, $\kappa=10^{-3}$).}
\label{fig_IQZE}
\end{figure}

This is explicitly demonstrated in figure~\ref{fig_IQZE}a) where we show results for $I_1$ as a function of $\varepsilon_1$ around the two-qubit resonance condition and for different values of $\varepsilon_2$. The 2qb-peak in the current through one qubit clearly moves as one varies the level position in the other, while the 1qb resonance remains unaltered (not shown) upon changing $\varepsilon_2$. We can also note that as the difference $|\Omega_i-\omega_r|$ increases, the effective couplings given by equations~(\ref{JeffZ}) and (\ref{JeffXZ}) decrease and therefore the induced qubit-qubit interaction is turned off. Experiments along these lines, with individual addressing of the QDs,  have been recently reported for transport through carbon-nanotube quantum dots, where non-local control mediated by a photon cavity (in the classical limit) has been demonstrated \cite{delbecq13}. Thus we expect that an experimental test of our prediction in figure~\ref{fig_IQZE}a) is within reach.

An even more interesting possibility is to non-locally manipulate the qubit-qubit induced interaction by tuning the \emph{dissipative} coupling of one of the qubits with its fermionic reservoirs. A strong coupling to the right reservoir in, say, qubit 2 induces a transport version of the quantum Zeno effect which tends to freeze the dynamics of the second qubit by effectively localizing the charge in the left dot of the DQD2, with $\langle\sigma_z^2\rangle\rightarrow1$. We demonstrate this effect in figure~\ref{fig_IQZE}b) where the current through the first DQD as a function of $\Gamma_{R,2}$ is shown for the two-qubit resonance condition $\varepsilon_1=-\varepsilon_2$, with $\Gamma_{\beta,1}=\Gamma_{L,2}=\Gamma$. It is observed there that for fixed qubits parameters, the current through DQD1 is strongly reduced by increasing merely the rate $\Gamma_{R,2}$ of the second DQD. We can reinforce the interpretation of this results by recalling the effective qubit-qubit interaction: for very large $\Gamma_{R,2}$ one can replace the operators of the second qubit by the corresponding mean value; then the effective coupling constants for the first qubit are also frozen and results in a smaller effective coupling. A strong coupling to a dissipative bath could also lead to charge localization and hence to effectively destroy the qubit-qubit interaction mediated by photons.
\begin{figure}[t]
\centering
    \includegraphics[width=0.85\columnwidth]{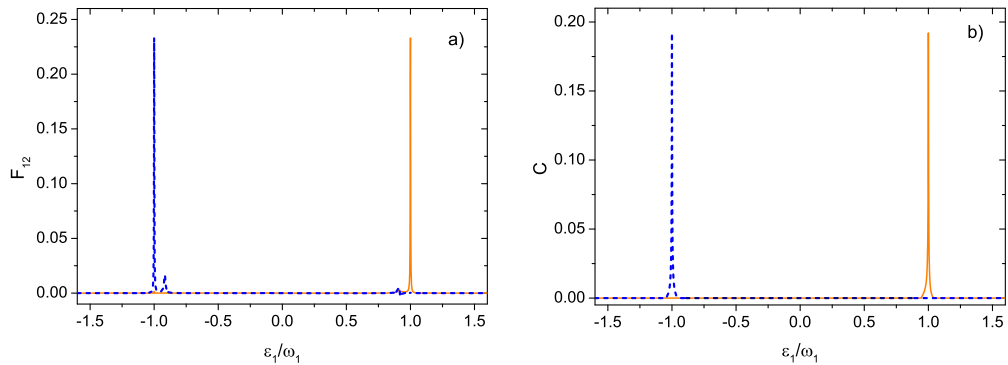}
   \caption{a) Correlators for the current across both qubits, $F_{21}$, and b) Concurrence, for $\varepsilon_{2}=-1$ (solid line) and $\varepsilon_{2}=1$ (dashed line). Same parameters as in figure~\ref{fig_I2qb}.}
\label{fig_S21C}
\end{figure}

To check more critically the presence of non-local correlations mediated by the microwave resonator, we study how nontrivial noise correlations develop. The Fano factor for the DQD1 shows the same qualitative behaviour exhibited in the single-qubit case (for symmetric rates with the reservoirs), with sub-Poissonian regions around all resonances of the problem, see figure~\ref{fig_I2qb}b).

\begin{figure}[t]
\centering
   \includegraphics[width=0.33\columnwidth,angle=-90]{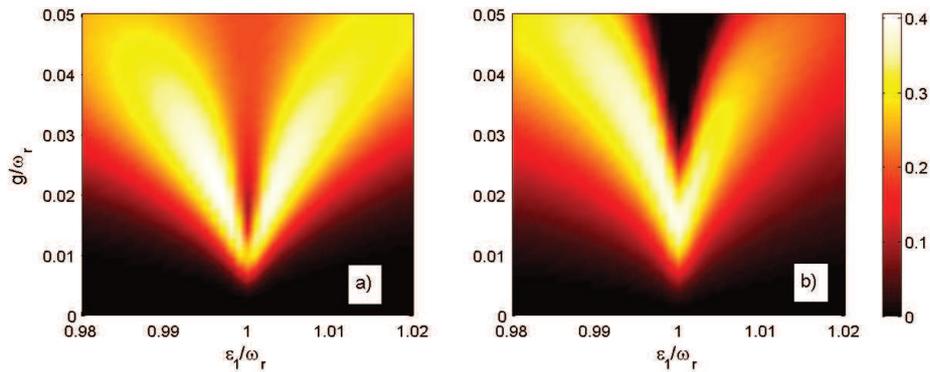}
   \caption{Colormap of a) cross-correlations $F_{12}$ and b) Concurrence around the two-qubit resonance as a function of the level detuning $\varepsilon_{1}$ and the coupling parameter with the resonator, $g$. Rest of the parameters (in units of $\omega_r$): $\varepsilon_2=-1$, $t=0.2$, $\Gamma=10^{-3}$, $\kappa=10^{-3}$.}
     \label{fig_MF12C}
\end{figure}

More crucially, the cross-correlations between separate currents through both DQDs, $F_{12}$, develop sharp resonances at the qubit-qubit resonance $\Omega_1=\Omega_2$, figure~\ref{fig_S21C}a). Apart from these clear resonances, other small features signal finite microwave resonator occupations which lead to non-zero correlations. As the coupling with the resonator increases, such features, and more generally the overall behaviour as a function of level detuning, can become rather intricate. Figure~\ref{fig_MF12C}a) shows the cross-correlations for increasing $g$ in the region around the 2qb resonance. This figure reveals that the peak emerged around this resonance splits as the qubits-resonator coupling becomes larger. At the same time, the resonances become broader such that the function $F_{12}(\varepsilon_{1},g)$ develops a two-lobe structure. As we shall show in the next Section, this characteristic structure signals the formation of Bell states between both qubits and hence the development of non-local entanglement.

\section{Qubit-qubit entanglement}\label{entanglement}

So far we have demonstrated that transport exhibits signatures of the induced interaction between the DQDs due to the common coupling to photons in the microwave resonator. Here, we go a step further an explicitly demonstrate that this common coupling can generate entanglement.
In particular, we show that qubit-qubit entanglement under \emph{nonequilibrium} conditions can be generated by virtual photons. For quantifying the nonequilibrium entanglement we make use of the Concurrence \cite{wootters}, a measure of entanglement defined by means of the density matrix of the system in the computational basis. We calculate the Concurrence of the steady state $\hat{P}\rho^{\mathrm{stat}}$, which corresponds to the projection of the stationary density matrix onto the two-qubits subspace with a proper normalization \cite{lambert07}, and trace out the states of the bosonic mode.

Numerical results for the Concurrence, $C$, considering the same interdot tunnelling amplitude in both qubits, $t_i=t$, and symmetric electron-photon coupling $g_{i}=g$ are shown in figure~\ref{fig_S21C}b) for two different level detunings in the second qubit. For the typical value of the coupling $g=0.008$ used here, $C$ shows sharp features in the 2qb resonance, $\varepsilon_{1}=-\varepsilon_{2}$.

In figure~\ref{fig_MF12C}b) we show the detail of Concurrence in the region of the 2qb resonance, as a function of $\varepsilon_1$ and the coupling strength to the microwave resonator $g$, for $\varepsilon_2=-1$. Here we find that, in the same way as the cross-correlators $F_{12}$, the peak exhibited by the Concurrence around resonance splits and develops a two-lobe structure as the coupling $g$ becomes larger. The similarity between these two quantities shows that current cross-correlations in the above configuration constitute an indicator of non-local qubit-qubit entanglement.

\begin{figure}
\centering
   \includegraphics[width=0.33\columnwidth,angle=-90]{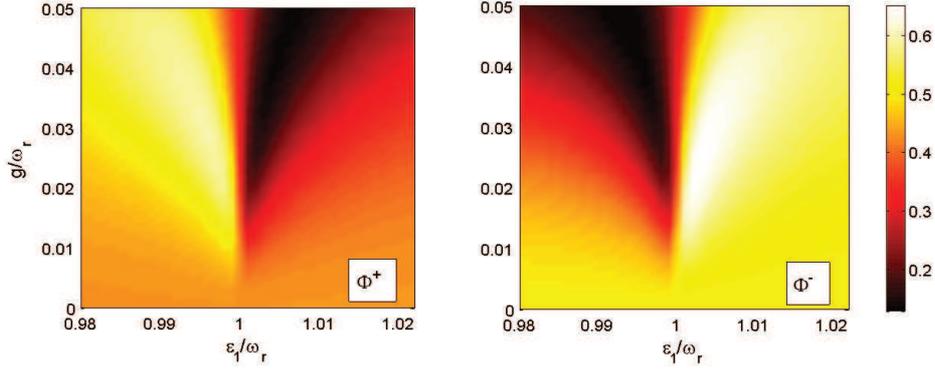}
   \caption{Stationary occupation probability of the Bell state $|\phi^{+}\rangle$ (left) and $|\phi^{-}\rangle$ (right) around the two-qubit resonance. Same axes and parameters as in figure~\ref{fig_MF12C}.
\label{fig_Mphim}}
\end{figure}

Although, to the best of our knowledge, a formal proof connecting noise cross-correlations and a finite steady-state concurrence  does not exist, the previous interpretation is supported by an analysis of the steady state populations of the system. If the analysis is done in terms of the populations in the local basis (e.g. $|\alpha_1,\alpha_2\rangle$, with $\alpha=L,R$) the double-peaked structure of figures \ref{fig_MF12C}a) and b) is hard to explain, since all local populations exhibit just a single peak around resonance.  However, considering the stationary populations in the Bell basis of maximally entangled states, a different picture arises.  Figure~\ref{fig_Mphim} shows the population of the Bell states $|\Phi^{\pm}\rangle$ \cite{popescu}, which written in terms of the occupation of the $L/R$ dots of each DQD read:
\begin{eqnarray}
  |\Phi^\pm \rangle = \frac{1}{\sqrt{2}}\left(|R_1,R_2\rangle\pm|L_1,L_2\rangle\right).
\end{eqnarray}
The occupation probability of these two states show a double peak structure as $g$ becomes larger. Importantly, these peaks occur asymmetrically such that each Bell state has maximum occupation on either side of the resonance. The two-lobe structure of both the Concurrence and the cross-correlations thus correspond to the two maxima of the $|\Phi^{\pm}\rangle$ Bell state populations.
The remaining Bell states
\begin{eqnarray}
  |\Psi^\pm \rangle = \frac{1}{\sqrt{2}}\left(|R_1,L_2\rangle\pm|L_1,R_2\rangle\right)
\end{eqnarray}
just show single peaks centered on resonance and presumably do not greatly contribute to the overall form of the current cross-correlations.

\section{Asymmetric coupling to the microwave resonator $g_1\neq g_2$.}\label{asymmetric}

Finally, we also explore the effect of asymmetric values of the electron-photon coupling strengths for each qubit, $g_1\neq g_2$. Experimentally, this asymmetry can be achieved by changing both the capacitive coupling of each DQD to the microwave resonator $C^c_{i}$ as well as the capacitance of each DQD to ground $C^g_i$, as the couplings scale as $g_i\sim C^c_{i}/(C^c_{i}+C^g_i)$ \cite{childress04}. Our motivation here is to explore the possibility of detecting the interaction-induced shifts directly in transport. Further motivation comes from \cite{lopez12}, which theoretically proposed the use of inhomogeneous coupling between two-level systems and a single quantized mode to generate and control multipartite entangled states.

\begin{figure}
\centering
      \includegraphics[width=0.3\columnwidth,angle=-90]{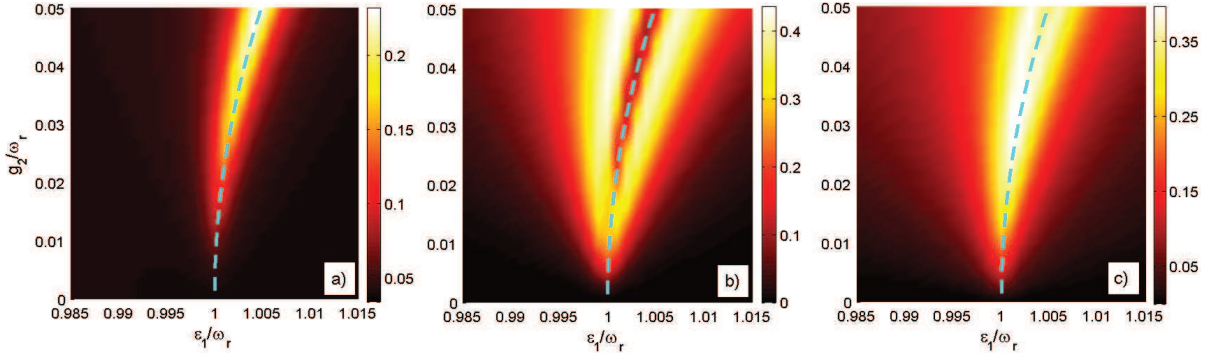}
   \caption{Colormap of a) Current across the first qubit, b)cross-correlations $F_{12}$ and b) Concurrence as a function of $\varepsilon_{1}$ and $g_{2}$, for fixed $g_{1}=0.008$, and around the two-qubit resonance. The dashed line indicate the renormalized two-qubit resonance $\Omega_{1,\mathrm{eff}}=\Omega_{2,\mathrm{eff}}$. Rest of the parameters as in figure~\ref{fig_MF12C}}
   \label{fig_mapIFCeD}
\end{figure}

The current as function of $\varepsilon_{1}$ and $g_2$ is shown in figure~\ref{fig_mapIFCeD}a) for the region around $\varepsilon_1\approx-\varepsilon_2$. For increasing $g_2$, the position of the resonance is shifted with respect to the initial value for $g_1=g_2$. As expected, this can be understood by means of the renormalization of the intra-dot tunnelling coupling $t_{i,\mathrm{eff}}$ in the effective Hamiltonian of equation~(\ref{Heff}). This renormalization leads in turn to a change in the frequency of the qubits as $\Omega_{i,\mathrm{eff}}=\sqrt{\varepsilon_i^2+4t_{i,\mathrm{eff}}^2}$. Therefore, the current shows a dispersive shift at values of $\varepsilon_1$ accordingly to the new, effective qubit-qubit resonance condition given by $\Omega_{1,\mathrm{eff}}=\Omega_{2,\mathrm{eff}}$. The dispersive shift obtained with the full numerics agrees with the one given by the effective Hamiltonian, represented by the dashed line in figure~\ref{fig_mapIFCeD}. Measurements along these lines would constitute further proof of resonator-induced interaction between qubits. The same dispersive shift is also observed in the shot noise cross-correlations, figure~\ref{fig_mapIFCeD}b), where again, the 2qb resonance in $F_{12}$ splits for large enough coupling.

Finally we present the same analysis for the Concurrence in figure~\ref{fig_mapIFCeD}c). Apart from the shift, we can notice that, in general, the Concurrence has larger values in comparison to the case with $g_1=g_2$, indicating that the asymmetry between the coupling parameters of each qubit with the bosonic mode makes the qubit-qubit entanglement to be more robust.

\section{Conclusions}\label{conclusions}

We studied theoretically photon-mediated transport and the generation of steady state correlations between two open charge qubits defined in spatially-separated double quantum dots which are coupled to a common transmission line resonator. Our results demonstrate that the qubits are entangled due to the indirect coupling induced by photons in the microwave resonator. Considering that each qubit is open to electronic reservoirs, we have analysed their transport properties and found that they reveal the qubit-qubit interaction. In particular, we calculated the zero-frequency shot noise and the current cross-correlations as a function of the level detuning of one of the qubits, and observed the presence of different resonant features in the regions where the qubit enters in resonance with the photon as well as with the other qubit. In the examples we studied here, the quantum correlations involved in the transport of charge and which are responsible of the signal in the cross-correlations, yield in a finite value for the Concurrence when the qubits interact due to off-resonant photons. Therefore, we propose that measurements of current correlations could be used as a possible method for detecting entanglement and, in general, qubit-qubit interactions mediated by the microwave resonator. This proposal is motivated also in the context of recent experimental achievements demonstrating the coupling of semiconductor QDs to microwave resonators \cite{frey11,delbecq12,frey12,frey12b,toida12}.

The model presented here constitute a step further in the study of this kind of hybrid systems, which can be relatively easily extended to several qubits. In general, this system led us to explore the interplay between coherent interactions, entanglement and the effect of dissipation and noise. Moreover, our model can also be applied to charge qubits defined by Cooper-pair boxes or to systems in which the quantum dots are coupled to a nanoelectromechanical resonator.

Finally, we can also mention that our results point to interesting future work in which the DQDs parameters are systematically modified such that the degree of qubit-qubit entanglement can be improved and even controlled.

\ack
We are grateful to N. Lambert, F. Nori and P. Samuelsson for interesting discussions and for letting us know about their works \cite{samuelsson13,lambert13} previous to submission to the arXiv. We would like to thank T. Kontos, E. Cota and E. Solano for helpful comments. This work was supported by the European Commission (STREP PICC), the Alexander von Humboldt Foundation, the DAAD and DFG grant numbers BR 1528/7-1, 1528/8-1, SFB 910 and GRK 1558 and the Spanish MINECO through grant numbers FIS2009-08744 and FIS2012-33521.

\appendix
\section{Effective Hamiltonian}\label{appendix}
We start with the Hamiltonian for the isolated electronic system and resonator:
$H_\mathrm{2} \equiv H_0 + V$ with $H_0 = H_{\mathrm{el}} + H_{\mathrm{res}}$ and $V= H_{\mathrm{e-res}}$.  We then move to the representation for the qubit operators
\begin{equation}
  \tilde{\sigma}_z^i
  =
  \frac{\epsilon_i}{\Omega_i} \sigma_z^i
  +
  \frac{2t_i}{\Omega_i} \sigma_x^i
  ;\quad
    \tilde{\sigma}_x^i
  =
  \frac{\epsilon_i}{\Omega_i} \sigma_x^i
  -
  \frac{2t_i}{\Omega_i} \sigma_z^i
  \label{eHam2}
  ,
\end{equation}
with frequency $\Omega_i = \sqrt{\varepsilon_i^2 + 4 t_i^2}$, which diagonalizes the electronic Hamiltonian $H_{\mathrm{el}}$.

We assume that both DQDs are singly occupied and then use second-order perturbation theory to find the matrix elements between states of the form $|s_1,s_2,0\rangle$ which has an empty cavity and qubits in states $s_1$ and $s_2$. In doing so we restrict intermediate excitations to states with just a single photon, $|s'_1,s'_2,1\rangle$.

This gives the effective Hamiltonian
\begin{eqnarray}
  H_{\mathrm{eff}}&=&
  \frac{1}{2} \sum_i \Omega_i \tilde{\sigma}_z^i
  -\frac{1}{\omega_r}\left(A_1 \tilde{\sigma}_z^1 +A_2 \tilde{\sigma}_z^2\right)^2
  \nonumber\\
  && + \frac{1}{2}\sum_i B_i^2\left(
    \frac{1}{\Omega_i - \omega_r}\left(1+\tilde{\sigma}_z^i\right)
    -
    \frac{1}{\Omega_i+ \omega_r}\left(1-\tilde{\sigma}_z^i\right)\right) \nonumber \\
&& + \frac{1}{2} B_1 B_2 \sum_i\left(
    \frac{1}{\Omega_i - \omega_r}
    -
    \frac{1}{\Omega_i+ \omega_r} \right)\tilde{\sigma}_x^1\tilde{\sigma}_x^2 \nonumber\\
  &&
  -  \frac{1}{2}\sum_i A_i B_i \left(
    \frac{1}{\Omega_i - \omega_r}
    +
    \frac{1}{\Omega_i+ \omega_r} \right)\tilde{\sigma}_x^i \nonumber\\
  &&
  +\frac{1}{2} A_1 B_2
  \left(
    -\frac{2}{\omega_r}
    +\frac{1}{\Omega_2 - \omega_r}
    -
    \frac{1}{\Omega_2+ \omega_r}
   \right)
  \tilde{\sigma}_z^1 \tilde{\sigma_r}_x^2 \nonumber\\
  &&
  +\frac{1}{2} A_2 B_1
  \left(
    -\frac{2}{\omega_r}
    +\frac{1}{\Omega_1 - \omega_r}
    -
    \frac{1}{\Omega_1+ \omega_r}
  \right)
  \tilde{\sigma}_x^1 \tilde{\sigma}_z^2
  \label{eHam1}
\end{eqnarray}
with parameters
\begin{equation}
  A_i = \frac{g_i \varepsilon_i}{\Omega_i}
  ;\quad
  B_i = -\frac{2 g_i t_i}{\Omega_i}
  .
\end{equation}

Under the further assumptions that $|\Omega_i-\omega_r| \ll g$ (dispersive limit) and  $|\Omega_i-\omega_r| \ll \Omega_i+\omega_r$ (rotating wave approximation), we then obtain the effective Hamiltonian used in the main text:
\begin{eqnarray}
  H_\mathrm{eff}
  &=&
  \sum_i \frac{1}{2} \varepsilon_i \sigma_z^i + t_i \sigma_x^i
  + \frac{g_i^2 t_i}{\Omega_i(\Omega_i-\omega_r)} \sigma_x^i
  \nonumber\\
  &&
  +\sum_i \frac{g_1 g_2 t_i^2}{\Omega_i^2 (\Omega_{i}-\omega_r)}\sigma_z^1\sigma_z^2
  -\sum_{i\neq j} \frac{g_i g_i \varepsilon_i t_j^2}{\Omega_j^2 (\Omega_{j}-\omega_r)}\sigma_z^i\sigma_x^j
  .
\end{eqnarray}

\section*{References}

\end{document}